# Delay of light in an optical bottle resonator with nanoscale radius variation: dispersionless, broadband, and low-loss

M. Sumetsky

*OFS Laboratories, 19 Schoolhouse Road, Somerset, NJ  08873, USA*

[*]*sumetski@ofsoptics.com*

It is shown theoretically that an optical bottle resonator with a nanoscale radius variation can perform a multi-nanosecond long dispersionless delay of light in a nanometer-order bandwidth with minimal losses. Experimentally, a 3 mm long resonator with a 2.8 nm deep semi-parabolic radius variation is fabricated from a 19 micron radius silica fiber with a sub-angstrom precision. In excellent agreement with theory, the resonator exhibits the impedance-matched 2.58 ns (3 bytes) delay of 100 ps pulses with 0.44 dB/ns intrinsic loss. This is a miniature slow light delay line with the record large delay time, record small transmission loss, dispersion, and effective speed of light.

The speed of photons is always equal to the speed of light *c*. However, a light pulse propagating through an optical structure does not get from point $\mathbf{r}_a$ to point $\mathbf{r}_b$ in time $|\mathbf{r}_a - \mathbf{r}_b|/c$ since it can be absorbed and reemitted, reflected, trapped by a resonant state, travel through a curved waveguide, etc. Regardless of the propagation details the effective speed of light can be determined as $v = L/\tau$, where $\tau$ is the actual time of travel from $\mathbf{r}_a$ to $\mathbf{r}_b$ and $L \geq |\mathbf{r}_a - \mathbf{r}_b|$ is the size of the propagation region.

The intriguing problem is to identify the photonic structure with the smallest size $L$ that can perform the required delay $\tau$ of a pulse of width $\Delta\tau$ without distortion. The quest for such structures is central in the slow light research [1-6]. Beside general interest, these structures are of great importance for their potential key role as delay lines in optical computing and transformation of data on a chip. For this reason, the research efforts were targeted at the demonstration of a slow light delay line (SLDL) with the smallest dimensions for a given delay time and bandwidth and smallest possible attenuation and dispersion of pulses. Solving this problem is complicated by the fundamental delay-bandwidth product limitation which establishes the smallest possible dimensions of a photonic structure enabling the time delay $\tau$ of a pulse with the spectral width $\Delta\lambda$ [7, 8]. In addition, the major impediments for experimental solution of the problem are the yet insufficient fabrication precision of the modern photonic technologies and the attenuation of light in lithographically fabricated photonic structures [3-6, 9].

To arrive at the smallest dimensions, the SLDLs are usually engineered from chains of coupled microresonators fabricated lithographically. The microresonators can be created of microrings [3, 6] and can also be introduced in photonic crystals as microscopic perturbations of the periodic lattice [4, 5]. Due to bouncing of light in each microresonator, the propagation time increases and can be sufficiently large for long chains. In practice, the performance of these devices is limited by material and scattering losses, while the delay time strongly depends on the inter-resonator and resonator-waveguide coupling. For weaker coupling, light propagation along the delay line is slower and the delay time is larger, however, the transmission bandwidth proportionally decreases following the delay-bandwidth product dilemma [7, 8].

The state of the art fabrication precision of microscopic photonic elements is as small as a few nm [3]. However, this high precision is still not sufficient for the creation of the smallest possible SLDLs. Consider, e.g., a delay line consisting of coupled ring resonators with radius $r \sim 10$ μm. At telecommunication wavelength $\lambda \sim 1.5$ μm, the fluctuation $\delta r \sim 1$ nm of the resonator radius results in the spectral fluctuations $\delta\lambda \sim \lambda \delta r / r \sim 0.15$ nm. These fluctuations corrupt the SLDL performance because they are comparable to the characteristic spectral bandwidth of telecommunication pulses. Similar fabrication problem persists for other types of SLDL. In addition, in spite of the remarkable progress in fabrication of engineered miniature photonic SLDLs [5, 6], their characteristics suffer from significant attenuation caused by material and scattering losses. As the result of fabrication inaccuracy and material and scattering losses, the attenuation of miniature SLDL demonstrated to date, is very large and is measured in the range of 10-100 dB/ns [10]. Thus, the insufficient precision of the lithographically fabricated miniature photonic SLDL and attenuation of light in these structures call for alternative solutions.

This Letter proposes and demonstrates a different type of SLDL, which, in contrast to the miniature delay lines considered previously [1-6, 10, 11], is not based on the periodic and quasi-periodic photonic structures. It has the shape of an extremely elongated bottle resonator [12] with semi-parabolic radius variation (Fig. 1) [13]. It is found that, to arrive at the minimum possible axial speed of light within the bandwidth of several tenths of a nanometer at telecommunication wavelengths, the depth of the parabolic radius variation has to be as small as a few nm. The developed theory shows that a bottle resonator SLDL with nanoscale radius variation can be impedance-matched to the input/output waveguide and perform a multi-nanosecond delay of light at telecommunication wavelengths within a nanometer-order bandwidth having minimal losses and dispersion.

The key experimental result presented below is the actual demonstration of the bottle resonator SLDL with a breakthrough performance. To fabricate this resonator, the recently developed SNAP (Surface Nanoscale Axial Photonics) technology is employed. It enables the predetermined nanoscale modification of the optical fiber radius with the record sub-angstrom precision by annealing with a focused $CO_2$ laser beam [14, 15]. As a result, a 3 mm long resonator with a 2.7 nm deep semi-parabolic radius variation is fabricated from a 19 micron radius silica fiber with a sub-angstrom precision. In excellent agreement with theory, the resonator exhibits the impedance-matched and dispersionless 2.58 ns (3 bytes) delay of 100 ps pulses with 0.44 dB/ns intrinsic and 1.2 dB/ns full losses, which surpass the losses demonstrated to date for the miniature SLDL [3-6,10] by more than an order of magnitude.

To arrive at the smallest possible axial speed of light within the nanometer-order transmission bandwidth, we consider a bottle resonator with dramatically small effective radius variation $\Delta r(z)=r(z)-r_0$ of several nanometers only [16]. This small variation can completely confine the whispering gallery modes (WGMs) propagating along the fiber between turning points $z_{t1}(\lambda)$ and $z_{t2}(\lambda)$ [17]. In the cylindrical frame of reference $(z,\rho,\varphi)$, the coordinate dependence of the WGM field is separable and expressed as $\exp(im\varphi)\Lambda_n(\rho)\Psi(z)$ with orbital and radial quantum numbers *m* and *n*. The *z*-dependence of the WGM propagating along the fiber is described by the one-dimensional Schrödinger equation [17]

$$\Psi_{zz}+(E(\lambda)-V(z))\Psi=0 \qquad (1)$$

with propagation constant $\beta(\lambda,z)=\sqrt{E(\lambda)-V(z)}$. In Eq. (1) the energy $E(\lambda)$ and potential $V(z)$ are determined as

$$E(\lambda)=-\kappa^2\frac{\Delta\lambda}{\lambda_{res}}, \quad V(z)=-\kappa^2\frac{\Delta r(z)}{r_0}, \quad \kappa=\frac{2^{3/2}\pi n_r}{\lambda_{res}}, \tag{2}$$

where $\Delta\lambda=\lambda-\lambda_{res}-i\gamma$ is the wavelength variation near a resonance $\lambda_{res}$, $n_r$ is the refractive index of the fiber, and $\gamma$ determines the attenuation of light in the fiber. The transmission amplitude $S(\lambda,z)$ and group delay of the resonator coupled to the input/output microfiber at contact point $z=z_c$ are [14]:

$$S(\lambda,z_c)=S_0-\frac{i|C|^2 G(\lambda,z_c,z_c)}{1+DG(\lambda,z_c,z_c)}, \quad \tau(\lambda,z_c)=\frac{\lambda^2}{2\pi c}\text{Im}\left(\frac{\partial \ln S(\lambda,z_c)}{\partial \lambda}\right) \tag{3}$$

where $S_0$ is the out-of-resonance amplitude, $C$ and $D$ are the bottle resonator/microfiber coupling constants, and $G(\lambda,z_1,z_2)$ is the Green's function of the wave equation. In the semiclassical approximation,

$$G(\lambda,z,z)=\frac{\cos(\varphi(\lambda,z_{t1},z)+\chi_1)\cos(\varphi(\lambda,z,z_{t2})+\chi_2)}{\beta(\lambda,z_1)\sin(\varphi(\lambda,z_{t1},z_{t2})+\chi_1+\chi_2)}, \quad \varphi(\lambda,z_1,z)=\int_{z_1}^{z}\beta(\lambda,z)dz. \tag{4}$$

where $\chi_i$ are the phase increments near the turning points $z_{ti}$, while the turning points are found from equation $\beta(\lambda,z_{ti})=0$. Eqs. (3) and (4) fully describe the bottle resonator SLDL. Averaging the group delay $\tau(\lambda,z)$ found from Eqs. (2)-(4) over the local period of these oscillations yields the average group delay, which, for relatively small coupling loss [11], coincides with the classical time of the roundtrip propagation along the bottle resonator

$$\overline{\tau(\lambda)}=\frac{\lambda_{res}^2}{\pi c}\int_{z_{t1}}^{z_{t2}}\frac{\partial\beta(\lambda,z)}{\partial\lambda}dz \tag{5}$$

A light pulse launched from the input microfiber into the bottle resonator at contact point $z_c$ slowly propagates along the resonator axis in both directions and returns back after reflecting from turning points $z_{t1}$ and $z_{t2}$. Generally, after completing the roundtrip between one of the turning points and the contact point, the pulse does not fully return back into the microfiber output and is partly reflected back into the resonator. The reflection at $z_c$ determines the impedance mismatch between the input/output microfiber and resonator and causes bouncing of the pulse between turning points with decreasing amplitude. In the stationary formulation described by Eq. (3) and (4), oscillations of the pulse correspond to oscillations of the transmission amplitude and group delay as a function of wavelength. Suppression of these oscillations leads to the condition of impedance matching. Under this condition, the pulse is fully transmitted from the microfiber input into the bottle resonator and, after completing the roundtrip along the resonator axis $z$, it is fully transmitted back into the microfiber output.

It is shown here that, for practical applications, it is possible and sufficient to solve the impedance matching problem for the bottle resonator SLDL locally. It is found that the oscillations of transmission amplitude vanish in the vicinity of wavelength $\lambda_i$ at a microfiber position $z_i$ close to the turning point $z_{t1}$ under the conditions (see Supplemental Material [18]):

$$\mathrm{Im}(S_0)=0, \ |C|^2=2S_0\mathrm{Im}(D), \tag{6}$$

$$\beta(\lambda_i, z_i)=\mathrm{Im}(D), \tag{7}$$

$$\mathrm{Im}(D)/\mathrm{Re}(D)=\tan(\varphi(\lambda_i, z_{t1}, z_i)+\chi_1). \tag{8}$$

Eqs. (6) are similar to the condition of lossless resonator/microfiber coupling [14], while Eqs. (7) and (8) determine the relationship between the contact point $z_i$, wavelength $\lambda_i$, and coupling parameter $D$.

Generally, propagation of an optical pulse though the bottle resonator SLDL is dispersive. To avoid dispersion, the eigenfrequencies of the bottle resonator should be locally equidistant, which is typical for large and smooth quantum wells $V(z)$ away from their bottom. However, the slowest axial speed of light corresponds to the bottom of quantum well. To arrive at the dispersionless propagation with the smallest possible speed, the shape of the bottle resonator in this region should have the equidistance frequency spectrum. Ignoring the slow $\lambda$-dependence of $\chi_1$ and $\chi_2$, this condition is satisfied for the bottle resonator with semi-parabolic radius variation (Fig. 1(b)):

$$\Delta r(z) = \begin{cases} \Delta r_0 - \dfrac{z^2}{2R} & \text{for } 0 < z < L = (2R\Delta r_0)^{1/2}, \\ 0 & \text{elsewhere,} \end{cases} \quad (9)$$

where $R$ is the axial curvature of the bottle resonator and $L$ is its length. The classical wavelength-independent time delay in this structure found from Eq. (5) is

$$\tau = \frac{\pi n_r L}{c}\sqrt{\frac{r_0}{2\Delta r_0}}. \quad (10)$$

An illustration of a semi-parabolic bottle SLDL is given in Fig. 1. The resonator parameters in this figure are chosen to model the experiment below (fiber radius $r_0 = 19\ \mu$m, semi-parabola height $\Delta r_0 = 2.8$ nm, resonator length $L = 3$ mm, speed of light $c = 3 \cdot 10^8$ m/s, and fiber refractive index $n_r = 1.46$). For these parameters, Eq. (10) yields the delay time $\tau = 2.67$ ns and effective speed of light $L/\tau = c/267$. Inside the bottle resonator near its edge $z = z_{t1}$ the propagation constant is independent of the axial coordinate $z$. In this region, to arrive at the impedance matching condition at wavelength $\lambda_1$ we determine the imaginary part of the coupling parameter $D$ from Eq. (7) and then the coordinate $z_1$ for arbitrary $\text{Re}(D)$ from Eq. (8). Next, we determine $|C|^2$ from Eq. (6). Fig. 2(a) and (b) are the surface plots of the resonance amplitude and group delay distributions as a function of wavelength and distance along the bottle resonator in the vicinity of $z_{t1}$ for the determined coupling parameters. The spectral profiles crossing the impedance matched point $(\lambda_1, z_1)$ are shown in Fig. 2 (c) and (d). The amplitude and group delay ripples vanish at $(\lambda_1, z_1)$ and are relatively small in the neighborhood of this point. Fig. 2(e) shows the time domain propagation of a 100 ps pulse through the constructed SLDL (the amplitude spectrum of the pulse is depicted in Fig. 2(c)). It is seen that the spurious temporal ripples at the output are remarkably small (less than 8.5% in magnitude (Fig. 3(b)) and thus less than 0.7% in intensity) and the FWHM pulse broadening in Fig. 3(b) is negligible. The average delay time found from the spectrum in Fig. 2(d) and from the time domain calculation in Fig. 2(e) is in excellent agreement with the value 2.67 ns found from Eq. (10).

Once a semi-parabolic bottle SLDL with delay time $\tau$, pulse width $\Delta\tau$, and spurious temporal ripples magnitude $\varepsilon$ is determined, an SLDL enabling the dispersionless delay of a pulse with a different temporal width $\Delta\tau_s = s\Delta\tau$ and similar performance can be simply constructed by rescaling the height of the parabola $\Delta r_0$ and its length $L$ to

$$\Delta r_s = s^{-1} \Delta r_0, \quad L_s = s^{-1/2} L. \tag{11}$$

From Eq. (10), these transformations do not change the delay time $\tau$. The magnitude of spurious ripples of the new delay line remains the same as well (see Supplemental Material [18]). Thus, once the set of Eqs. (6)-(8) is satisfied, the semi-parabolic bottle resonator can be impedance-matched to an input/output waveguide and perform a multi-nanosecond delay at telecommunication wavelengths within a nanometer bandwidth having negligible losses and dispersion.

Experimentally, the bottle resonator was created at the 19 µm radius optical fiber by annealing with a focused $CO_2$ laser beam. The speed of the beam was varied to ensure the required semi-parabolic profile of $\Delta r(z)$, which was characterized as follows. First, the surface plot of transmission amplitude as a function of wavelength $\lambda$ and microfiber position $z$ was measured by scanning the microfiber waist along the resonator [19, 20]. Then, the introduced radius variation $\Delta r(z)$ was calculated from the measured spectrum following theory [14] (Fig. 3(b)). The resonator had the length of 3 mm and depth of $\Delta r(z)$ of 8 nm. The parabolic part of $\Delta r(z)$ with the depth 2.8 nm and equidistant eigenfrequencies was introduced to ensure the dispersionless propagation of 100 ps pulses with the slowest-possible speed near the bottom of quantum well $V(z)$. The deviation of the parabolic part of $\Delta r(z)$ from the exact semi-parabola was less than 0.9 angstrom.

As compared to the theoretical model presented above, the experimentally realized bottle resonator SLDL is designed to have two sets of wavelengths and contact points, which determine regions with suppressed oscillations of the group delay and transmission amplitude. These sets, $(\lambda_1, z_1)$, and $(\lambda_2, z_2)$ (Fig. 4(a), (b)) correspond to the same microfiber/resonator coupling parameters $C$ and $D$, which gives us the opportunity to realize a miniature SLDL with a breakthrough performance as well as to determine the intrinsic loss of this device. At contact point $z_1$, the vicinity of wavelength $\lambda_1$ (Fig. 4(a),(b) and (c),(d)) corresponds to the propagation of light near the top of quantum well $V(z)$. This case is used below as a reference. At contact point $z_2$, the vicinity of wavelength $\lambda_2$ (Fig. 4(a),(b) and (e),(f)) corresponds to the slowest propagation in the parabolic part of quantum well $V(z)$, the case of our main interest.

Figures 4(g) and (h) illustrate the time-dependent propagation of a 100 ps Gaussian pulse calculated from the measured spectra shown in Fig. 4(c), (d) and (e), (f), respectively. The average group delays in Fig. 4(d) and (f) are in excellent agreement with the delay times 1.17 ns and 2.58 ns in Fig. 4(g) and (h) showing that the delay at the parabolic part of the bottle near wavelength $\lambda_1$ is more than two times greater than near $\lambda_2$. Comparison of the average transmission amplitudes in Fig. 4(c) and (e) and the corresponding delay times determines the intrinsic loss of the demonstrated device equal to 0.44 dB/ns. It is suggested that this loss is primarily caused by contamination of the resonator surface and can be significantly decreased in the clean room environment.

Since the performed optimization included only the taper translation and omitted tuning of the resonator and microfiber profile near the contact point, the optimal coupling was achieved only approximately. For this reason, the out-of-resonance coupling losses ~ 2 dB (irrelevant to intrinsic losses and, thus, independent of the delay time) were introduced. However, even in this case, the total insertion loss of the 2.58 ns (3 bytes) delay line has the impressive record value of 3 dB, i.e., 1.12 dB/ns, as compared to 10-100 dB/ns losses previously demonstrated for miniature delay lines [3-6,10]. The spurious temporal ripples in Fig. 4(h) are remarkably small (less than 11% in magnitude and thus less than 1.2% in intensity) and the FWHM pulse broadening is negligible (less than 3%, four times smaller than the pulse broadening for significantly smaller delay in Fig. 4(g)). In addition, the effective speed of light in this SLDL is $c/258$, the record small for the engineered slow light photonic structures [3-6].

This demonstration presents a solution to the central problem of the slow light research – creation of a miniature delay line with a breakthrough performance. In addition, it emphasizes the flexibility of the SNAP platform as a fruitful source for exciting fundamental and applied studies. Similar approach will allow creating a variety of photonic structures that precisely imitate one-dimensional quantum mechanical structures described by the potential $V(z)$ under interest. This includes potential structures that can be used for investigation of tunneling and time delay [21, 22, 23], Anderson localization [24, 25], localized states in continuum [26, 27], etc. This also includes intriguing opportunities for creating photonic microdevices for filtering, switching, lasing, delay of light, and sensing with the unprecedented high precision and low loss.

The author is grateful to Y. Dulashko for assisting in the experiments and to D. J. DiGiovanni and V. Mikhailov for helpful discussions.

**Figure captions**

Figure 1

(a) – Illustration of an optical bottle resonator delay line. Light is coupled into the resonator from a transverse waveguide (microfiber) and experiences whispering gallery mode propagation along the resonator surface. (b) – Semi-parabolic variation of a bottle resonator radius used in the numerical simulations.

Figure 2.

(a) and (b) – Surface plots of the transmission amplitude and group delay near the edge $z=z_{t1}$ of the bottle resonator calculated with the coupling parameters defined in the text. (c) and (d) – the transmission amplitude and group delay spectra at the coupling point $z_c=z_1$ shown in (a) and (b), respectively. (e) – The output signal amplitude (solid line) calculated for the input 100 ps pulse (dashed line). The pulse spectrum is determined by the bold line in (c).

Figure 3.

(a) – Illustration of an optical bottle resonator delay line. (b) – Experimentally measured surface plot the transmission amplitude, which was used to determine the bottle resonator radius variation (bold line).

Figure 4.

(a) and (b) – Surface plots of the transmission amplitude and group delay near the edge $z=z_{t1}$ of the fabricated bottle resonator measured after the optimization of coupling parameters by translation of the microfiber with respect to the resonator. (c) and (d) – the transmission amplitude and group delay spectra at the coupling point $z_c=z_1$ corresponding to minimum spectral oscillations in (a) and (b), respectively. (e) and (f) – the same as (c) and (d) but at the coupling point $z_c=z_2$, which corresponds to the semi-parabolic part of the potential $V(z)$. (g) and (h) – the output signal amplitudes (solid line) calculated for the input 100 ps pulse (dashed line) from the spectra measured at points $z_1$ and $z_2$ with the pulse spectrum determined by the bold line in (c) and (e), respectively.

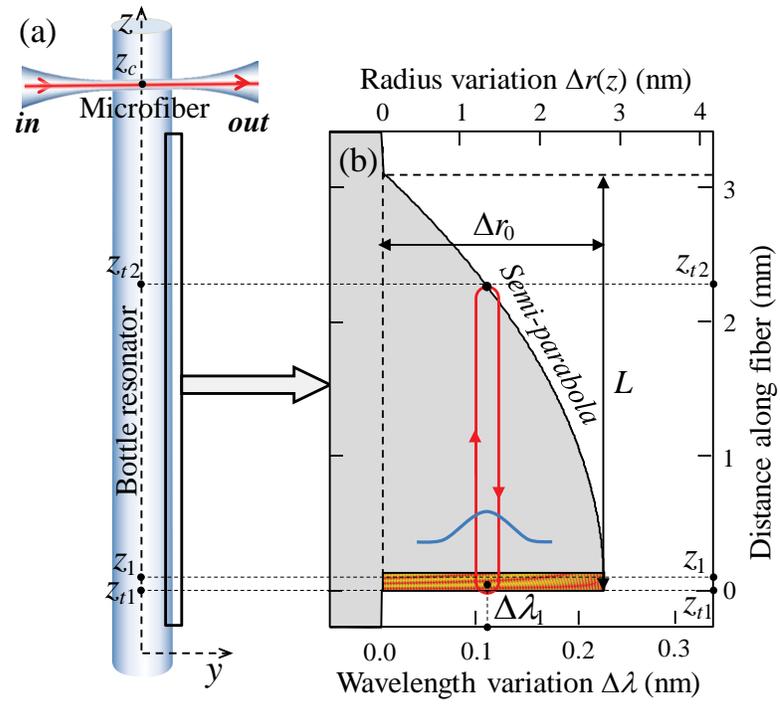

Figure 1.

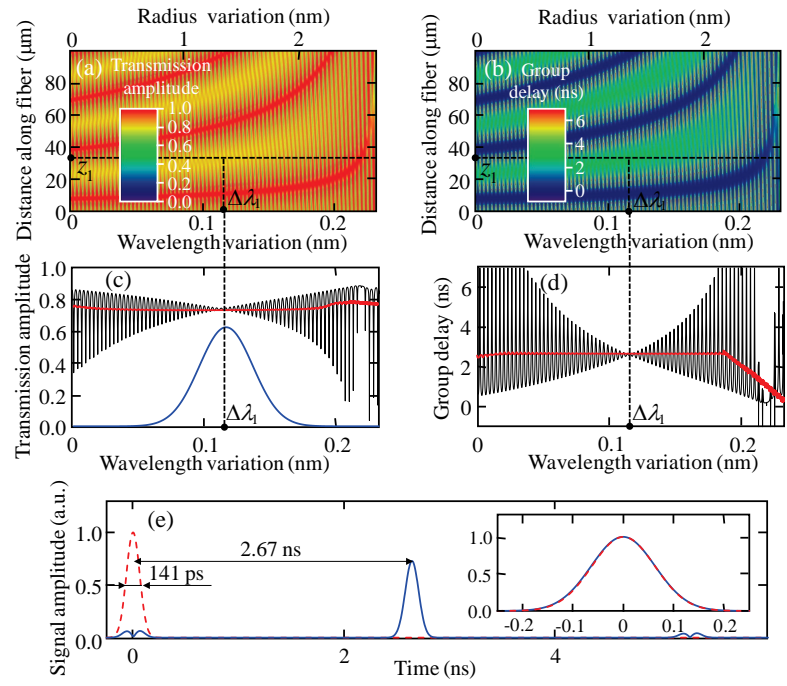

Figure 2.

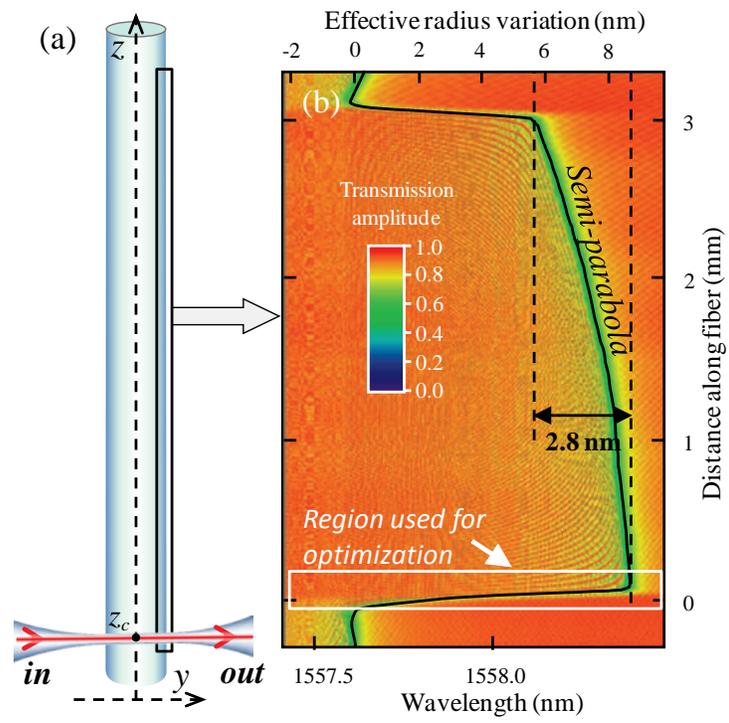

Figure 3.

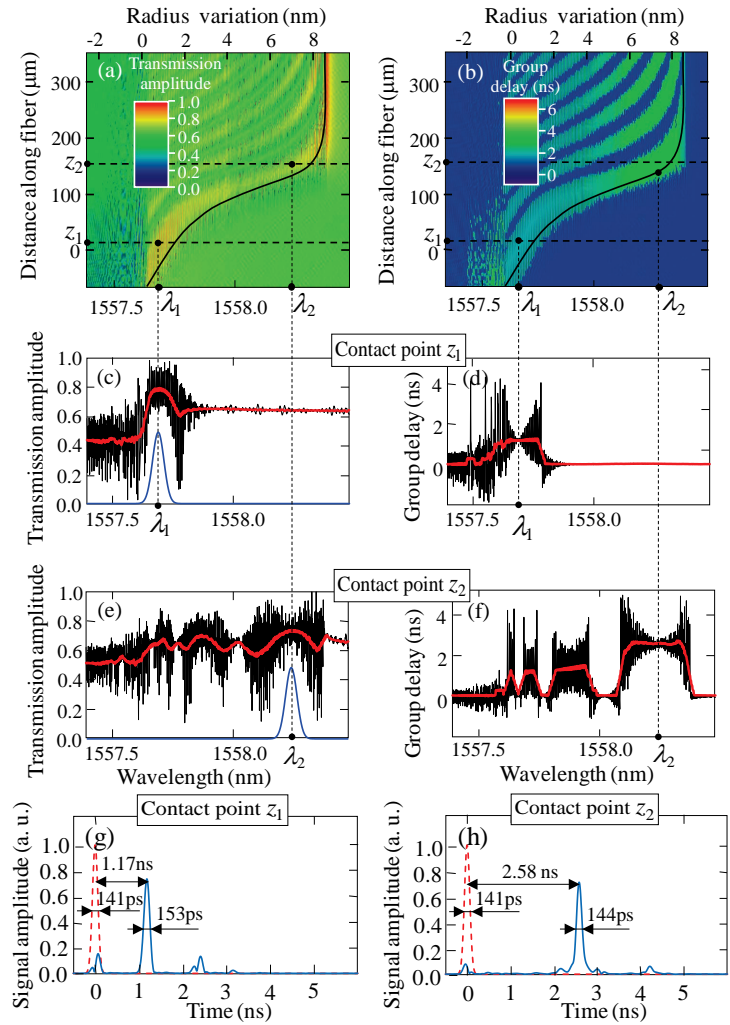

Figure 4.

## Supplemental Material

**Derivation of the impedance-matching conditions**

The semiclassical Green's function is given by Eq. (4). The transmission amplitude of a bottle resonator is defined by Eqs. (3) and (4) and exhibits strong oscillations as a function of wavelength. To find the condition of vanishing oscillations we write

$$|S(\lambda,z_c)|^2 = \left| S_0 - \frac{i|C|^2 G(\lambda,z_c,z_c)}{1+DG(\lambda,z_c,z_c)} \right|^2 = K^2 \quad \text{(S1)}$$

where $K$ is a constant. Taking into account that $G(\lambda,z_c,z_c)$ is real, Eq. (S1) is transformed to

$$\begin{aligned}
&\{[\text{Re}(S_0 D)]^2 + [\text{Im}(S_0 D) - |C|^2]^2 - K^2|D|^2\} G^2(\lambda,z_c,z_c) \\
&+ 2\{\text{Re}(S_0)\text{Re}(S_0 D) - K^2 \text{Re}(D) - \text{Im}(S_0)[\text{Im}(S_0 D) - |C|^2]\} G(\lambda,z_c,z_c) \\
&+ \{[\text{Re}(S_0)]^2 + [\text{Im}(S_0)]^2 - K^2\} = 0.
\end{aligned} \quad \text{(S2)}$$

Since Eq. (S2) should be valid for all values of $G(\lambda,z_c,z_c)$, three terms in curly brackets in this equation should be equal to zero. Simplifying the obtained equations yields Eq. (6) of the main text.

To determine the conditions when the oscillations of the group delay vanish, we represent the group delay from Eq. (3) as

$$\tau(\lambda,z_c) = \frac{i\lambda^2 |C|^2}{2\pi c(1+DG(\lambda,z_c,z_c))[S_0 + (DS_0 - i|C|^2)G(\lambda,z_c,z_c)]} \frac{\partial G(\lambda,z_c,z_c)}{\partial \lambda}. \quad \text{(S3)}$$

For briefness, we omit the dependencies on $\lambda$ and $z$ and define:

$$\varphi_{t12} = \varphi(\lambda,z_{t1},z_{t2}) + \chi_1 + \chi_2, \quad \varphi_1 = \varphi(\lambda,z_{t1},z_c) + \chi_1. \quad \text{(S4)}$$

It is assumed that the contact point $z_c$ is situated close to the turning point $z_{t1}$ (Fig. 1). Then $|\varphi_1| \ll \varphi_{t12}$ and $\varphi_1$ is a much slower function of wavelength than $\varphi_{t12}$. Taking this into account, Eq. (S3) is transformed to

$$\frac{\partial G}{\partial \lambda} = \frac{|C|^2 \cos^2(\varphi_1)}{\beta(\lambda,z_c)(a_1 \cos(\varphi_{t12}) + b_1 \sin(\varphi_{t12}))(a_2 \cos(\varphi_{t12}) + b_2 \sin(\varphi_{t12}))} \int_{z_{t1}}^{z_{t2}} \frac{\partial \beta(\lambda,z)}{\partial \lambda} dz \quad \text{(S4)}$$

$$\begin{aligned}
a_1 &= 1 + \frac{D}{\beta(\lambda,z_c)} \cos(\varphi_1)\sin(\varphi_1), \quad b_1 = -\frac{D}{\beta(\lambda,z_c)} \cos^2(\varphi_1), \\
a_2 &= S_0 + \frac{DS_0 - i|C|^2}{\beta(\lambda,z_c)} \cos(\varphi_1)\sin(\varphi_1), \quad b_2 = -\frac{DS_0 - i|C|^2}{\beta(\lambda,z_c)} \cos^2(\varphi_1).
\end{aligned} \quad \text{(S5)}$$

Fast oscillations of the denominator in Eq. (S4) are eliminated if

$$a_1 = \pm ib_1 \quad \text{and} \quad a_2 = \mp ib_2. \quad \text{(S6)}$$

Substitution of Eq. (S5) into Eq. (S6) yields Eqs. (6), (7), and (8).

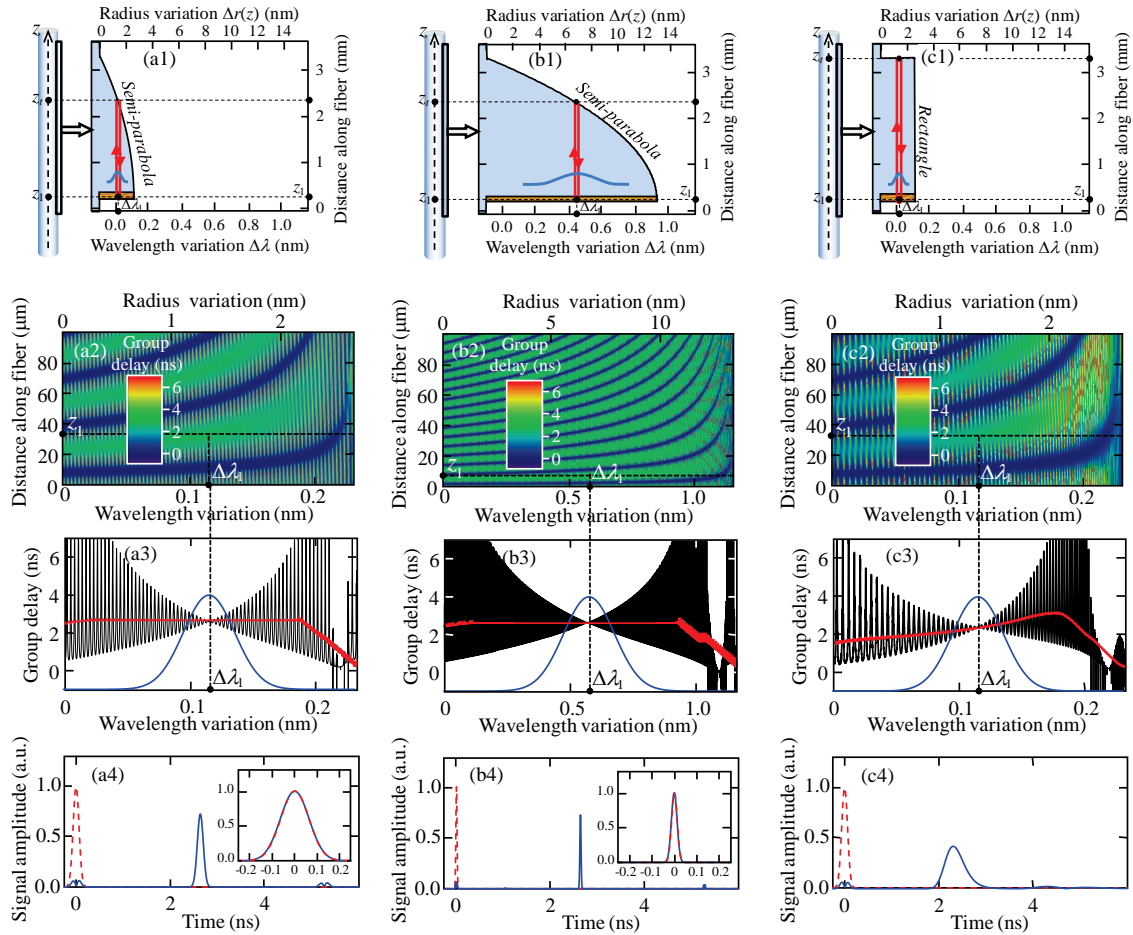

Fig. S1. Comparison of the semi-parabolic SLDL considered in the main text, (a1)-(a4), with the semi-parabolic SLDL designed to propagate 5 times narrower pulses, (b1)-(b4), and with the rectangular SLDL having the same radius variation depth, (c1)-(c4). (a1),(b1), (c1) – Radius variations of the semi-parabolic, rescaled semi-parabolic, and rectangular bottle resonators. (a2), (b2), (c2) – surface plots of the group delay spectra near the edge of these resonators. (a3), (b3), (c3) – The group delay spectra of these resonators at the point of contact with the input/output microfiber. (a4), (b4), (c4) – The input and output signal amplitudes as a function of time for these resonators.

**Rescaling of parabolic bottle resonator SLDL. Parabolic vs. rectangular bottle resonator SLDL.**

Transformation of the shape of semi-parabolic bottle resonator with Eq. (11) allows us to construct a SLDL which delays pulses of different widths with the same performance (i.e., the same small spurious ripples and negligible pulse dispersion). As an example, Fig. S1(a1)-(a4) and (b1)-(b4) compare the SLDL designed in the main text to delay 100 ps pulses with that designed to propagate 5 time narrower (20 ps) pulses with the delay time and same quality (i.e., for $s=1/5$ in Eq. (11) of the main text). The period of rapid oscillations of the group delay as a function of wavelength is determined by the delay time and, for this reason, is the same in Fig. S1(a2), (a3) and (b2), (b3). The envelope of oscillations in Fig. S1(b3) is obtained from the envelope of oscillations in Fig. S1(a3) by the linear expansion along the wavelength axis with the expansion coefficient of the pulse spectral width, $1/s=5$. For this reason, the magnitude of the spurious ripples in the output signal temporal dependence shown in Fig. S1(a4) and (b4) does not change and the pulse dispersion is negligible.

It is also instructive to compare the parabolic SLDL with the rectangular one with the same depth of radius variation shown in Fig. S1(c1). It is seen that the non-uniformity of the period of fast spectral oscillations, which can be detected from Fig. S1(c3), leads to strong dispersion of the output pulse in Fig. S1(c4).